\documentclass[a4paper,fleqn,usenatbib]{mnras}
\usepackage{amssymb,amsmath,graphicx,psfrag,mathtools}
\usepackage[T1]{fontenc} 
\usepackage{ae,aecompl} 
\usepackage{url}
\title[Why is ISO 1I/2017 U1 Rocky and Very Prolate?]{Why is Interstellar Object 1I/2017 U1 (`Oumuamua) Rocky, Tumbling and Very Prolate?}
\author[J. I. Katz]{
J. I. Katz,$^{1}$\thanks{E-mail katz@wuphys.wustl.edu} 
\\
$^{1}$Department of Physics and McDonnell Center for the Space Sciences,
Washington University, St. Louis, Mo. 63130 USA 
}
\date{Accepted XXX.  Received YYY; in original form ZZZ} 
\pubyear{2017} 
\date{\today}
\begin{document} 
\label{firstpage} 
\pagerange{\pageref{firstpage}--\pageref{lastpage}} 
\maketitle 
\begin{abstract}
The recently discovered first interstellar object 1I/2017 U1 (`Oumuamua)
has brightness that varies by a factor of 10, a range greater than that of
any Solar System asteroid, a spectrum characteristic of Type D asteroids,
and no evidence of evaporating volatiles, contrary to expectation for
exo-Oort clouds.  This object was the first example of the proposed
``Jurads'', objects depleted in volatiles and ejected from planetary systems
during the post-main sequence evolution of their parent star.  I suggest
that heating by the star's giant stage fluidized a precursor cometary object
as volatiles escaped, causing it to assume the Jacobi ellipsoidal shape of a
self-gravitating incompressible liquid.  The collision that produced the
inferred tumbling motion must have occurred thousands of years after the
formation of 1I/2017 U1 `Oumuamua.  Jacobi ellipsoids have a unique relation
among rotation rate, density and axial ratio.  The inferred axial ratio
$\gtrapprox 5$ implies a lower bound on the density of 1.6 g/cm$^3$,
excluding an icy interior unless it is almost entirely frozen CO$_2$.  This
object is the first Jurad to be discovered and may be related to accreting
objects that pollute white dwarf atmospheres and may make Soft Gamma
Repeaters.
\end{abstract}
\section{Introduction}
The recently discovered interstellar object (ISO) 1I/2017 U1 (`Oumuamua)
is remarkable for its hyperbolic Solar orbit \citep{Me17,Ma17}, its deep
brightness modulation (a factor of ten, greater than that of any Solar
System asteroid) \citep{Me17,Fr17,D17} and the absence (with tight upper
bounds) of evaporating volatiles.  The light curve is not strictly periodic
but rather indicates tumbling \citep{Fr17,D17}.  Although interstellar
objects expelled from exo-planetary systems have been predicted for many
years \citep{HZ17}, they have been predicted to be overwhelmingly
cometary, evaporating volatiles and emitting dust during close approaches to
the Sun (1I/2017 U1 `Oumuamua approached within 0.25 AU \citep{Me16}).
Thus, while the discovery of an interstellar object was not unexpected, its
properties require explanation.

The brightness modulation of 1I/2017 U1 `Oumuamua implies an extraordinary
axial ratio and indicates a novel origin.  The very slight ($\lesssim 100$
erg/cm$^3$) strength required to support the less prolate nonequilibrium
shapes of rotating Solar System bodies of similar size indicates that this
interstellar object had an extraordinary history during which its strength
was reduced to essentially zero.  Its unexpected absence of volatiles is
also remarkable, and these two facts point towards a hypothesis of its
origin and history.

Here I discuss the possible origin of 1I/2017 U1 `Oumuamua in the post-main
sequence stage of a planetary system \citep{HZ17,RAVQB17}.  The increasing
luminosity of a post-main sequence star will evaporate volatiles from its
cometary satellites.  Even those in Oort-cloud like orbits will pass close
enough to the star to be heated because those orbits are nearly parabolic,
with small periastrons.  The rapid mass loss of
white dwarf formation or the essentially instantaneous mass loss of neutron
star formation in a core-collapse supernova will unbind many of the small
bodies of the exoplanetary system.  It will also put massive exoplanets on
elliptic orbits in which they will interact gravitationally with each other
and with the small bodies, expelling some and putting others on the low
angular momentum orbits required to explain the ``pollution'' of white dwarf
atmospheres \citep{JY14,F16,X17,VSG17,Mu17} by heavier elements.  Additional
mechanisms of expulsion of small bodies operate in binary systems
\citep{S17}.

One consequence of heating is the loss of volatile ices, turning cometary
nuclei into asteroidal bodies without further volatiles to evaporate,
explaining the absence of volatiles in 1I/2017 U1 `Oumuamua.  Evaporating
volatiles fluidize beds of refractory particles.  There is evidence for this
(at the lower levels of heating occurring in Solar System comets) in images
obtained by the Rosetta spacecraft of Comet 67P/Churyumov-Gerasimenko
\citep{R17}.  Regions near the center of the cometary nucleus, near the
minimum of its gravitational potential, are covered by fine material, in
contrast to the rough craggy appearance of the rest of the body.  This is
likely the result of flow of fluidized particulates to the minimum of the
gravitational potential.  A comet heated by the intense radiation of a red
giant star may be entirely fluidized and the residual refractory material
assume the shape, a Maclaurin spheroid or, if it has more angular momentum,
a Jacobi ellipsoid, of a rotating self-gravitating incompressible liquid.

The purpose of this paper is to investigate the consequences of this
explanation of the unprecedentedly prolate shape of 1I/2017 U1 `Oumuamua.
This hypothesis implies a lower density limit that likely excludes the
possibility \citep{Fi17} of an icy interior protected from Solar heating by
an insulating mantle.  Finally, the inference that exoplanetary systems may
be disrupted by the evolution of their stars and that white dwarf pollution
may be caused by single bodies as massive as $\sim 10^{22}$ g \citep{X17},
or possibly even more massive \citep{J09}, is consistent with the
explanation of Soft Gamma Repeater outbursts as the result of accretion of
such minor-planet sized bodies onto neutron stars \citep{KTU94}.
\section{Geometry}
Jacobi ellipsoids are the equilibrium state of uniformly rotating
strengthless homogeneous incompressible fluids with too much angular
momentum to be (oblate) Maclaurin spheroids.  Even were their specific
angular momentum to be increased, strengthless Jacobi ellipsoids will not
break up (spinning mass off their extremities) provided they have time to
relax to their equilibrium shape.  Extremely elongated Jacobi ellipsoids may
be unstable to ``pear-shaped'' (actually, more like egg-shaped) $\ell = 3$,
``dumb-bell'' $\ell = 4$ and higher instabilities \citep{EHS82,HE82} but
the nonlinear development of these instabilities is not understood so here I
assume Jacobi ellipsoids.

The ten-fold modulation of the brightness of 1I/2017 U1 `Oumuamua would
imply, in a na\"{i}ve model that assumes a prolate spheroid, Lambert's Law
reflectivity with uniform albedo and oppositional geometry (the actual Solar
angle during photometric observations was about 20$^\circ$), an axial ratio
of 10:1 if viewed in the plane of rotation, and greater for other viewing
angles \citep{Me17}.  However, Lambert's Law is not valid for Solar System
asteroids.  Modeling that assumes 1I/2017 U1 `Oumuamua is described by the
scattering properties of known Solar System asteroids \citep{Fr17,D17} sets
a lower bound on the axial ratio of about 5:1.  No upper bound on the axial
ratio can be obtained because the angle between the line of sight and the
rotation axis is unknown, but a lower bound of about 65$^\circ$ can be set
on the angle \citep{D17}.  Assuming a plausible upper bound on the density
of 3 g/cm$^3$ leads to an upper bound on the axial ratio of about 8
(Sec.~\ref{density}), from which a somewhat more stringent lower bound on
the angle may be found.

\begin{figure}
\centering
\includegraphics[width=0.99\columnwidth]{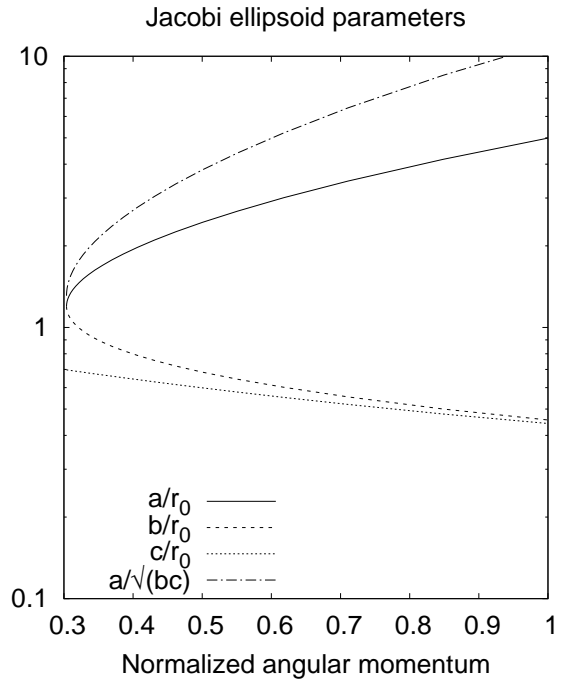}
\caption{\label{jacobi1} Semi-principal axes $a$, $b$ and $c$ of Jacobi
ellipsoids as a function of the normalized angular momentum ${\cal L} \equiv
L/\sqrt{GM^3r_0}$ where $r_0 \equiv (abc)^{1/3}$ \citep{C69}.  At higher
angular momentum ${\cal L} \gtrsim 0.5$ $b \to c$.  For lower angular
momenta (${\cal L} < 0.3$, outside the range of the figure) $a = b > c$ and
the body is an oblate Maclaurin spheroid.  $a/\sqrt{bc}$ is the axial ratio
used in Fig.~\ref{jacobi3}.}
\end{figure}

Uniformly rotating (friction very quickly enforces uniform rotation)
self-gravitating incompressible fluids of uniform density are oblate 
Maclaurin spheriods (with no modulation of their scattered light) if their
angular momentum is low, or triaxial Jacobi ellipsoids if they have higher
angular momentum \citep{C69,T78}.  For large angular momenta the two smaller
semi-axes $b$ and $c$ of the Jacobi ellipsoids converge, as shown in
Fig.~\ref{jacobi1}, and the ellipsoid approaches a prolate spheroid.
\section{Density}
\label{density}
Jacobi ellipsoids have a unique relation among their semi-axes, density and
rotation rate that can be used to establish bounds on their density from the
measured rotation rate and the axial ratio inferred from their light curve.
Alternatively, assumed bounds on their density can be used to constrain
their geometry.

\begin{figure}
\centering
\includegraphics[width=0.99\columnwidth]{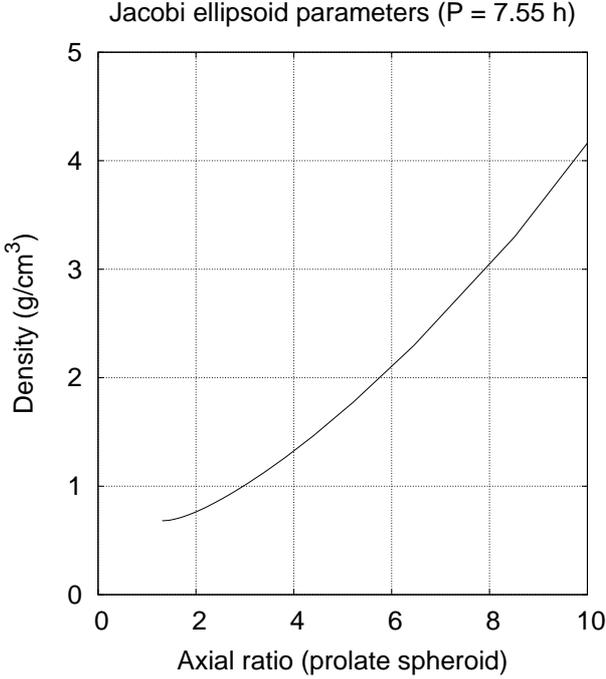}
\caption{\label{jacobi3} Relation between density and axial ratio (defined
as $a/\sqrt{bc}$) of a Jacobi ellipsoid with rotation period 7.55 h
\citep{C69}.  The empirical lower bound of 5 on the axial ratio implies a
lower bound of 1.6 g/cm$^3$ on the density.}
\end{figure}

Several slightly different values of the rotation rate have been reported
\citep{Me17,Fr17,D17,B17}.  Adopting a period of 7.55 h \citep{D17}, roughly
the mean of these values, leads to the relation between axial ratio and
density shown in Fig.~\ref{jacobi3}.  The lower bound on the axial ratio of
about 5 fitted \citep{D17,Fr17} to the light curve implies a minimum density
of 1.6 g/cm$^3$.  This is inconsistent with all plausible icy materials
except CO$_2$, whose density is close to this value; even the density of
CO$_2$ is inconsistent unless we happen to lie almost exactly in the plane
of rotation, the axial ratio is less than inferred from the light curve or
the rotation period is significantly longer than the adopted value.  This
argues against the hypothesis \citep{Fi17} that an icy core is protected
against heating and evaporation by an insulating nonvolatile mantle.
\section{Tumbling: The Evolutionary History}
The tumbling motion of 1I/2017 U1 `Oumuamua \citep{Fr17,D17} implies that it
is not rotating exactly around its axis of greatest moment of inertia.
Tumbling requires at least minimal elastic strength to maintain.   It must
have been the result of collision after the epoch of fluidization because
any viscous, plastic or frictional flow would lead to very rapid relaxation
to periodic rotation about the axis of greatest moment of inertia (the
lowest energy state).  The required strength is difficult to estimate (we
have only the crudest hint of the rotational dynamics), but is less than the
extremely small central pressure $p$.  For high axial ratios $a \gg b
\approx c$ (Fig.~\ref{jacobi1}) and the ellipsoid can be approximated as an
infinite cylinder of radius $b$:
\begin{equation}
p \approx \pi G \rho^2 b^2 \sim 20\ \text{erg/cm}^3
\end{equation}
for a density $\rho \sim 3$ g/cm$^3$ and radius $b \sim 30$ m.

\citet{Fi17} indicate a thermal diffusivity $D \sim 10^{-4}$ cm$^2$/s.  This
implies a cooling time for an object of this size
\begin{equation}
t_{conduction} \sim {b^2 \over D} \sim 10^{11}\ \text{s}.
\end{equation}
In order for a collision to have produced tumbling it must have occurred
after 1I/2017 U1 `Oumuamua was no longer fluidized, either because the
volatiles were exhausted or because it cooled below volatilization
temperatures.  The red giant stages of stellar evolution are longer than
$t_{conduction}$ so a cometary body of these dimensions would have heated
throughout.

If 1I/2017 U1 `Oumuamua was completely fluidized and did not recover 
mechanical strength while still hot but depleted of volatiles, then at
least thousands of years (and possibly much longer) must have elapsed from
the end of the red giant stage of the parent star and intense heating to the
collision that made it tumble; 1I/2017 U1 `Oumuamua must have remained
in the fossil planetary system surrounding the former red giant, where the
density of solid bodies was high enough to make collision likely, for at
least that long before being expelled.

The dynamical processes that expel small bodies also put other small bodies
on collision orbits with their parent stars.  This is consistent both with
the inference that polluted white dwarfs accrete solid bodies for $\sim
10^9$ y and with the active lifetimes of Soft Gamma Repeaters of a few
thousand years after the supernov\ae\ in which their neutron stars were
born.
\section{Discussion}
The data contain rich implications that were not expected when interstellar
objects were first hypothesized:
\begin{enumerate}
\item Some of the light curves of 1I/2017 U1 `Oumuamua \citep{Me17} are
asymmetric about their minima, rising more slowly than they fell.  This may
be explained as a shadowing effect if the rotation is prograde with
respect to its orbital motion (an accident with 50\% probability of
occurrence for an interstellar object).
\item The extraordinary and unanticipated axial ratio of 1I/2017 U1
`Oumuamua, combined with the absence of volatiles, indicates its origin in
the exoplanetary system of a star that had passed through the red giant
stage.
\item The combination of the rotation period with the inferred large axial
ratio sets a lower bound on the density of 1I/2017 U1 `Oumuamua that
excludes an icy composition, even one under a non-volatile crust that might
be consistent with the observed absence of a coma.
\item The inference that 1I/2017 U1 `Oumuamua is tumbling sets further
constraints on its history: it likely remained in a region of comparatively
high density of solid objects for at least thousands of years after it
formed.
\item The discovery of an interstellar interloper confirms expectations from
the study of ``polluted'' white dwarfs, and suggests that analogous
processes may produce Soft Gamma Repeater outbursts.
\end{enumerate}
\section*{Acknowledgements}
I thank B. Hansen and B. Zuckerman for useful discussions.
\bibliography{iso}
\bsp 
\label{lastpage} 
\end{document}